

\input phyzzx
%
\catcode`\@=11 
\def\NEWrefmark#1{\step@ver{{\;#1}}}
\catcode`\@=12 

\def\square{\kern1pt\vbox{\hrule height 1.2pt\hbox{\vrule width
   1.2pt\hskip 3pt
 \vbox{\vskip 6pt}\hskip 3pt\vrule width 0.6pt}\hrule height 0.6pt}\kern1pt}
\def\lb{\lbrack\!\lbrack}
\def\rb{\rbrack\!\rbrack}

\def\bra#1{\langle #1 |}
\def\ket#1{| #1 \rangle}

\def\A{{\cal A}}
\def\B{{\cal B}}
\def\C{{\cal C}}
\def\D{{\cal D}}
\def\H{\widehat{\cal H}}

\def\W{{\cal W}}

\def\L{{\cal L}}
\def\M{{\cal M}}

\def\O{{\cal O}}
\def\P{{\cal P}}

\def\X{{\cal X}}
\def\Y{{\cal Y}}
\def\Z{{\cal Z}}

\def\V{{\cal V}}

\def\p{\partial}

\def\wh{\widehat}

\def\B{{\cal B}}
\def\W{{\cal W}}
\def\P{{\cal P}}
\def\V{{\cal V}}
\def\O{{\cal O}}

\def\p{\partial}


\def\define#1#2\par{\def#1{\Ref#1{#2}\edef#1{\noexpand\refmark{#1}}}}
\def\con#1#2\noc{\let\?=\Ref\let\<=\refmark\let\Ref=\REFS
         \let\refmark=\undefined#1\let\Ref=\REFSCON#2
         \let\Ref=\?\let\refmark=\<\refsend}

\let\refmark=\NEWrefmark

\define\zwiebachlong{B. Zwiebach, `Closed string field fheory: Quantum
action and the Batalin-Vilkovisky master equation', Nucl. Phys {\bf B390}
(1993) 33, hep-th/9206084.}

\define\spivak{M. Spivak, ``Calculus on Manifolds'', Benjamin/Cummings
publishing (1965).}

\define\hatazwiebach{H. Hata and B. Zwiebach, `Developing the covariant
Batalin-Vilkovisky approach to string theory',
Ann. Phys. {\bf 229} (1994) 177, hep-th/9301097.}

\define\schwarz{A. Schwarz, `Geometry of
Batalin-Vilkovisky quantization', UC
Davis preprint, hep-th/9205088, July 1992. }

\define\senzwiebachgauge{A. Sen and B. Zwiebach, `A note on gauge
transformations in Batalin-Vilkovisky theory',
Phys. Lett. {\bf B320} (1994) 29,
hep-th/9309027.}

\define\wittenab{E. Witten, `A note on the antibracket formalism', Mod.
Phys. Lett. {\bf A5} (1990) 487.}

\define\getzler{E. Getzler, `Batalin-Vilkovisky algebras and two-dimensional
topological field theories', MIT Math preprint, hep-th/9212043.}

\define\lianzuckerman{B. H. Lian and G. Zuckerman,
``New Perspectives on the BRST-algebraic structure of string theory'',
hep-th/9211072, Yale preprint, November 1992.}

\define\penkavaschwarz{M. Penkava and A. Schwarz,
``On Some Algebraic Structures Arising in String Theory",
UC Davis preprint, UCD-92-03, hep-th/9212072.}

\define\senzwiebachtwo{A. Sen and B. Zwiebach, `Quantum background
independence of closed string field theory', MIT preorint CTP\#2244,
hep-th/9311009; to appear in Nucl. Phys. B}

\define\wittenunp{E. Witten, private communication, unpublished.}

\overfullrule=0pt
\baselineskip 16pt plus 1pt minus 1pt
\nopubblock
{}\hfill \vbox{\hbox{MIT-CTP-2346}
\hbox{hep-th/9408053}\hbox{August, 1994} }\break

\title{BACKGROUND INDEPENDENT ALGEBRAIC STRUCTURES}
\titlestyle{IN CLOSED STRING FIELD THEORY}

\author{Ashoke Sen \foot{E-mail  address: sen@theory.tifr.res.in,
sen@tifrvax.bitnet }  }
\address{Tata Institute of
Fundamental Research\break
Homi Bhabha Road, Bombay 400005, India}
\andauthor
{Barton Zwiebach \foot{E-mail address: zwiebach@irene.mit.edu,
zwiebach@mitlns.bitnet.\hfill\break Supported in part by D.O.E.
contract DE-AC02-76ER03069.}}
\address{Center for Theoretical Physics,
LNS and Department of Physics\break
MIT, Cambridge, Massachusetts 02139, U.S.A.}

\abstract
{We construct a  Batalin-Vilkovisky (BV) algebra on moduli spaces
of Riemann surfaces. This algebra is background independent in that
it makes no reference to a state space of a conformal field theory.
Conformal theories define a homomorphism of this algebra to the BV
algebra of string functionals. The construction begins with a
graded-commutative free associative algebra $\C$ built from the
vector space whose elements are orientable subspaces of moduli
spaces of punctured Riemann surfaces. The typical element here is a
surface with several connected components. The operation $\Delta$
of sewing two punctures with a full twist is shown to be an odd,
second order derivation that squares to zero. It follows that
$(\C, \Delta)$ is a Batalin-Vilkovisky algebra. We introduce the odd
operator $\delta = \partial + \hbar\Delta$, where $\partial$ is the
boundary operator. It is seen that $\delta^2=0$, and that consistent
closed string vertices define a cohomology class of $\delta$.
This cohomology class is used to construct a Lie algebra on a quotient
space of $\C$. This Lie algebra gives a manifestly background independent
description of a subalgebra of the closed string gauge algebra.}
\endpage

\chapter{Introduction and Summary}

At present the formulation of closed string field theory  requires two
choices. A choice of a set of string vertices, and a choice of a conformal
field theory representing a string background. It is now known that the use of
two different nearby sets of string vertices leads to the same string field
theory [\hatazwiebach]. Furthermore the use of two nearby conformal field
theories also leads to the same string field theory [\senzwiebachtwo]. This
latter property is called background independence. Since
a fundamental goal in
string theory is the writing of a manifestly background independent
formulation of the theory, investigation of
background independent structures
is an important endeavor.

In our earlier work [\senzwiebachtwo] we found  Riemann surfaces
analogs of the antibracket and the delta operator of Batalin-Vilkovisky (BV)
quantization. By making no reference to the state space of a conformal
theory, such objects define a background independent structure. We also
indicated that by including disconnected Riemann surfaces one would
obtain a complete BV algebra structure. The definition and main properties
of BV algebras have been considered in Refs.[\getzler,\penkavaschwarz,
\lianzuckerman]. The relevance of the Riemann surface BV algebra is
that a conformal background furnishes a natural map (homomorphism)
to the BV algebra of string functionals. Therefore, the Riemann surface BV
algebra is
a background independent object that underlies
the background dependent BV algebra of string fields.

There are three main points to the present work, and we discuss them now:

\noindent
(i) We give a detailed and precise description of the
relevant complex $\C$ of subspaces of moduli
spaces of disconnected punctured Riemann surfaces, and
introduce a graded associative algebra. We prove the existence of
a  BV algebra by introducing a delta operator $\Delta$ which is shown
to be a second order derivation that squares to zero. This part of the
work gives an economical derivation of the results of [\senzwiebachtwo]
and completes some of the details that were not given there.

\noindent
(ii) It was found in [\senzwiebachtwo] that the exponential of the
formal sum of closed string vertices $e^{\V/\hbar}$
defines an element of $\C$ that
is annihilated by the operator $\delta = \partial+\hbar\Delta$, where
$\partial$ is the boundary operator (picks the boundary of spaces of
surfaces). Moreover, $\delta^2 =0$. We were led to believe that the
problem of finding a consistent set of string vertices can be reformulated
as the problem of finding a cohomology class of $\delta$ on the complex $\C$.
We use here the earlier work of Ref.[\hatazwiebach] to show that given two
nearby sets of consistent string vertices $\V$ and $\V'$,
the difference
$e^{\V/\hbar} - e^{\V'/\hbar}$ is indeed $\delta$ trivial.
In obtaining this result
we had to include in the complex $\C$ some formal limits of spaces of
surfaces. This part of the work clarifies the geometrical basis of
the independence of string field theory on the choice of string vertices
and confirms the identification of string vertices with a cohomology class.

\noindent
(iii)  In Ref.[\senzwiebachgauge] we discussed the Lie algebra of gauge
transformations of a quantum field theory formulated in the BV approach.
These gauge transformations are built using the antibracket, the delta
operator and the master action. We build  a background independent
Lie algebra by using the geometrical antibracket, delta operator, and
the string vertices $\V$ (which are, except for the kinetic term, the
geometrical representative of the string action).
The Lie algebra is shown to be independent of the representative
$e^{\V/\hbar}$ for the cohomology class of $\delta$.
We show that the natural
map from spaces of surfaces to string functionals furnished by a conformal
background defines a homomorphism between the two Lie algebras.

There are some obvious and fundamental questions that we will not
address here. We have obtained a background independent BV algebra
and Lie algebra, and a homomorphism to background dependent BV algebras and
Lie algebras defined on string functionals.
We would like to know by how much the homomorphism fails to
be surjective. If the failure is small the
background independent algebraic structures discussed here capture much
of the string field theory algebraic structure. It is also not clear
how to use the background independent structures to give a manifestly
background independent construction of string field theory.
Witten [\wittenunp] has suggested that the
space of homomorphisms between the Riemann surface and string
field BV algebras
is a plausible candidate for the space of two-dimensional field theories.
This idea, and variations
thereof deserve concrete examination.

This paper is organized as follows. In sect.2 we construct the complex $\C$
and introduce the relevant associative algebra. In sect.3 we construct the
BV algebra on $\C$, discuss the $\partial$ and $\delta$ operators,
review the homomorphism to the BV algebra of string functionals,
and introduce the operations of contraction and Lie derivatives on $\C$.
In sect.4 we discuss cohomology of $\delta$ in
$\C$, changes of string vertices,
and the background independent Lie algebra built using the string vertices.

\chapter{The Associative Graded-Commutative Algebra}

Let us begin by defining the spaces we are going to work with.
We let $\M^g_n$ denote the moduli space of Riemann surfaces of genus $g$
and with $n$ punctures. The space $\P^g_n$ will denote the moduli
space of Riemann surfaces of genus $g$ and with $n$ punctures with
a chosen analytic coordinate at each puncture. The space $\P^g_n$
is infinite dimensional  (except when $n=0$) as infinite
number of parameters are needed to define coordinates around punctures.
The space $\P^g_n$ has the structure of a fiber bundle over
$\M^g_n$, with
a projection that consists of forgetting about the analytic coordinates
at the punctures.

In closed string field theory it is useful to introduce the space
$\wh\P^g_n$ which is also a space fibered over $\M^g_n$. This space is
obtained from $\P^g_n$ by a projection that forgets the phase of the
local coordinate at each puncture. These spaces are useful because
they admit globally defined sections that extend all the way to the
boundary of $\M^g_n$. This is not the case for $\P^g_n$.
The spaces $\wh\P^g_n$ are also infinite dimensional, except when $n=0$.

We will see that the main geometrical operation having to do
with the BV antibracket is the operation of twist-sewing. This
is a natural operation on $\wh\P$ where we do not have the phases
available to do sewing with a fixed sewing parameter. In fact,
an antibracket in $\P$ defined by twist-sewing
would be degenerate and thus unacceptable.
Therefore we will only
analyze the BV structure on $\wh\P$.
We will begin by setting up the vector space $\C$ where the algebra
is defined, and then introduce the dot product.

\section{The Vector Space}

We will be interested in finite
dimensional orientable subspaces of $\wh\P^g_n$ for all $g,n \geq 0$.
Those spaces will be called {\it basic spaces}, to distinguish them
from generalized spaces to be introduced later.
These subspaces may or
may not have boundaries, and may or may not be connected, but will be
taken to be smooth submanifolds. A single surface of some genus $g$ and
some number of punctures $n$, is  a
basic space of zero dimension; a one parameter family of surfaces is
a basic space of dimension one. The dimension of a basic space can
easily exceed that of the moduli space $\M^g_n$; in such case the
basic space must contain families of surfaces that give the same
underlying surface upon forgetting about the local coordinates at
the punctures. For $n=0$ the dimensionality of a basic space cannot
exceed that of $\M^g_0\,(\equiv 6g-6)$.

Let us now introduce the space
$\wh\P^{\,(g_1,\cdots g_r)}_{(n_1,\cdots n_r)}$
which will be defined as the cartesian product of a finite number of
decorated moduli spaces mentioned above. We take
$$\wh\P^{\,(g_1,\cdots g_r)}_{(n_1,\cdots n_r)}
\equiv \wh\P^{g_1}_{n_1}\times \ldots \times
\wh\P^{g_r}_{n_r}\,\,.\eqn\one$$
A point in this space is a collection of surfaces
$(\Sigma_1, \ldots \Sigma_r)$  where $\Sigma_i\in \wh\P^{g_i}_{n_i}$.
We will think of $(\Sigma_1, \ldots
\Sigma_r)$ as a single generalized Riemann
surface, that is a surface  with $r$ disconnected components.
The space $\wh\P^{\,(g_1,\cdots g_r)}_{(n_1,\cdots n_r)}$ can therefore be
thought of as a space of disconnected Riemann surfaces. Given a collection
$\A_{g_i, n_i}\subset\wh\P_{g_i, n_i}$ of basic spaces of surfaces, we
can easily define a subspace of disconnected surfaces. We introduce
the {\it product space}
$$\A^{g_1}_{n_1}\times\cdots \times \A^{g_r}_{n_r}\in
\wh\P^{\,(g_1,\cdots g_r)}_{(n_1,\cdots n_r)}\eqn\two$$

This subspace of disconnected surfaces must also
have its punctures labeled.\foot{By an abuse of notation
we shall often refer to the puncture
on a Riemann surface associated with a
subspace $\A$ of the moduli space as a puncture of the space $\A$.}
Each disconnected surface has $N=\sum n_k$
punctures and they must be labeled from $1$ to $N$.
To start with we just have $r$ connected
components, with the $i$-th component having its punctures labeled from
$1$ to $n_i$.
Let $P^{(i)}_k$ denote the $k$-th puncture
of the space $\A^{g_i}_{n_i}$. In the relabeled object this will become
the puncture $P_{k+ \sum_{j<i} n_j}$.
This defines the labeling of the punctures
in the disconnected surfaces.

\noindent
\underbar{Orientation.}
Let $[\A^{g_i}_{n_i}]$ denote, at any point of $\A^{g_i}_{n_i}$, an
ordered basis of tangent vectors to $\A^{g_i}_{n_i}\subset \wh\P^{g_i}_{n_i}$.
This globally defined basis of
ordered vectors defines the orientation of $\A^{g_i}_{n_i}$.
Let $\{\A^{g_i}_{n_i}\}$ denote the tangent vectors in
$\wh\P^{\,(g_1,\cdots g_r)}_{(n_1,\cdots n_r)}$
induced by the tangent vectors $[\A^{g_i}_{n_i}]$ of
$\A_{g_i, n_i}$. Then the orientation of $\A^{g_1}_{n_1}\times \cdots \,
\times \A^{g_r}_{n_r}$ is defined by the ordered set of
tangent vectors $[\{\A^{g_1}_{n_1}\}, \ldots
\{\A^{g_r}_{n_r}\}]$.

\noindent
\underbar{Symmetric Spaces.}
As usual with spaces of surfaces with labeled punctures there is
a useful notion of a symmetric space. A basic space of surfaces $\A$
is said to be {\it symmetric} if the space obtained by exchanging the labels
of any two labeled punctures is exactly the same as the original space.
It follows from our definition of a product space that the punctures are
labeled in a specific way and under the exchange of labels the space
is not invariant. In order to define a notion of a symmetric product space,
we now introduce
a symmetrized product,
where we sum over all possible
ways of labeling the punctures in the resulting disconnected surfaces.
We define
$$\lb \A^{g_1}_{n_1},
\cdots ,\A^{g_r}_{n_r}\rb\equiv {1\over n_1!\cdots n_r!}
\sum_{\sigma\in S_N}  {\bf P}_\sigma  (\A^{g_1}_{n_1}\times\cdots
\times \A^{g_r}_{n_r})\, ,\eqn\twopp$$
where the sum $\sum_{\sigma\in S_N}$ runs over all permutations
of $N$ labels, and ${\bf P}_\sigma$ denotes the operator that changes
$P_n \to P_{\sigma (n)}$.
The above sum should be understood to be a formal sum where
we add up spaces multiplied by real numbers. The factors multiplying
the sum in the right hand side have been introduced for later convenience.
By construction, the space $\lb\A^{g_1}_{n_1},\A^{g_2}_{n_2}\,,
\cdots\,\A^{g_r}_{n_r}\rb$
is left invariant under the exchange of any two labels of a pair
of labeled punctures. In such a symmetric space any fixed
label puncture must appear in every disconnected component
of the generalized surfaces except for those components which do not
carry any puncture. The orientation of the space $\lb \A^{g_1}_{n_1},
\cdots \A^{g_r}_{n_r}\rb$ is defined by the orientation of the various
terms appearing on the right hand side of Eqn.\twopp.

It follows from our definition of $\lb\cdots \rb$ in Eqn.\twopp, that the
basic space
$\lb\A\rb$ is symmetric even if $\A$ is not. Furthermore,
for a symmetric basic space $\A$ one
has $\lb \A \rb = \A$, by virtue of the normalization factor included
in \twopp. The same normalization factor guarantees that
$$\lb \A_{n_1}^{g_1} , \cdots ,  \A_{n_r}^{g_r} \rb =
\lb\,  \lb\A_{n_1}^{g_1}\rb  , \cdots , \, \lb\A_{n_r}^{g_r}\rb \rb\, ,
\eqn\normmult$$
for arbitrary basic spaces $\A_{n_i}^{g_i}\,$.
Let us consider the case
when all the basic spaces of surfaces appearing in Eqn.\twopp\ are
symmetric and have the property that any given surface
$\Sigma\in \A^{g_i}_{n_i}$ with labeled punctures appears in
$\A^{g_i}_{n_i}$ with unit weight. (We shall
define such spaces $\A^{g_i}_{n_i}$ to be symmetric basic spaces
with unit weight.)
Then the sum in the right hand side can be rewritten as
a sum over inequivalent splittings of the $N$ labels in groups of
$n_1,n_2,\cdots ,n_r$
labels. Each inequivalent splitting will appear in the sum
$n_1!n_2!\cdots n_r!$ times as identical terms by virtue of the
symmetry of the basic spaces. As a consequence the sum in the right
hand side can be written as a sum over inequivalent splittings,
each term being a space of surfaces
with unit weight.
This means that each different labeled generalized surface
will appear with unit weight. This simple fact will be useful to
understand the consistency of the normalization factors of some of the
equations we shall encounter later.

 Since we will always be considering oriented
spaces (in the sense of homology) it is clear
that we must have
$$\lb\cdots\,,\A^{g_i}_{n_i}\,, \A^{g_{i+1}}_{n_{i+1}},\cdots\rb\,
=\,(-)^{\A^{g_i}_{n_i}\A^{g_{i+1}}_{n_{i+1}}} \,\lb
\cdots ,\A^{g_{i+1}}_{n_{i+1}}\,,\,\A^{g_i}_{n_i},\cdots \rb \,,\eqn\three$$
where the symbol $\A^{g_i}_{n_i}$ in the
exponent denotes the dimension of the space $\A^{g_i}_{n_i}$ (this notation
will be used throughout this paper).
Indeed, under the exchange of the basic spaces $\A^{g_i}_{n_i}$ and
$\A^{g_{i+1}}_{n_{i+1}}$,
the orientation of the generalized space
$\lb\A^{g_1}_{n_1}, \ldots \A^{g_r}_{n_r}\rb$
picks up the indicated sign factor.  The reader
may note that the symmetric assignment of labels to the punctures is necessary
for the exchange property to hold. If this was not the case the spaces to
the left and to the right of \three\ would not agree as spaces with labeled
punctures.

\noindent
\underbar{Grading.}
We grade the basic spaces $\A^g_n$ by their dimensionality,
which, as mentioned above, is a priori unrelated to the dimensionality
of the moduli space $\M^g_n$. This $Z$ grading induces an obvious $Z_2$
grading according to whether the dimensionality is even or odd.
The $Z$ grading of  generalized subspaces is defined
by the sum of
the dimensionalities of the basic spaces entering the definition of the
generalized space. Their $Z_2$ grading is also induced by the $Z$ grading.

\noindent
\underbar{The Complex $\C$.}
We finally introduce the complex where the BV algebra will be
defined.
This is the complex $\C$, with the structure of a vector space, whose
elements, denoted as $X, Y, \cdots$ are formal sums of the form
$$X =  \sum a^{g_1\cdots g_r}_{n_1\cdots n_r}\,
\lb\,\A^{g_1}_{n_1}, \cdots \, \A^{g_r}_{n_r}\,\rb\,,\eqn\four$$
with $a^{g_1\cdots g_r}_{n_1\cdots n_r}$ a set of real numbers
($r\geq 0$).
This vector space $\C$ is extraordinarily large! Is is spanned by
subspaces of decorated, disconnected surfaces of all genus
and of all numbers of punctures. When we add subspaces of generalized
Riemann surfaces simplification is only possible when the genus and
number of punctures of all the basic spaces match, and, in addition,
all but one of the basic spaces actually coincide.
For example, the addition of
$\lb\A^{g_1}_{n_1},\A^{g_2}_{n_2}\rb$ and
$\lb\B^{g_3}_{n_3}\,,\B^{g_4}_{n_4}\rb$,
with all $g_i$ and $n_i$ different, cannot be simplified. Not even
$\lb\A^{g_1}_{n_1},\A^{g_2}_{n_2}\rb + \lb\B^{g_1}_{n_1}\,,\B^{g_2}_{n_2}\rb$
can be simplified in general. Nevertheless
$$\lb\A^{g_1}_{n_1},\A^{g_2}_{n_2}\rb\,
+\, \lb\B^{g_1}_{n_1}\,,\A^{g_2}_{n_2}\rb
=\lb\,\A^{g_1}_{n_1}\,+\,\B^{g_1}_{n_1}\,\, ,\A^{g_2}_{n_2}\rb \, .\eqn\five$$
In general we simply take
$$\lb\A^{g_1}_{n_1},\A^{g_2}_{n_2}\,, \cdots\, \rb
+\, \lb\B^{g_1}_{n_1}\,,\A^{g_2}_{n_2}\,\cdots\,\rb
=\lb\,\A^{g_1}_{n_1}\,+\,\B^{g_1}_{n_1}\,\, ,\A^{g_2}_{n_2}\,,\cdots \rb
\, ,\eqn\six$$
where all the basic spaces implied by the dots are the same, one by one,
in the three terms appearing in the equation.
The zero element $\bf 0$ in the vector space can be identified with
any generalized subspace $\lb\A^{g_1}_{n_1},\A^{g_2}_{n_2}\,, \cdots\, \rb$,
where one (or more) of the basic spaces is the empty set of surfaces.

Since the vector space $\C$ is spanned by symmetric elements we call
$\C$ the space of symmetric subspaces of direct
products of basic spaces. A general
element in $\C$ is symmetric in the sense that
every sector with a fixed numbers of punctures is.
The complex $\C$ is actually spanned by symmetrized products of
{\it symmetric} basic spaces. This is clear from eqn.\normmult,
where an arbitrary symmetrized product is rewritten as the symmetrized
product of a set of basic spaces that are symmetric.
Furthermore,
since each symmetric basic space can be expressed as linear combinations
of symmetric basic spaces with unit weight, it follows that the
complex $\C$ is spanned by symmetric products of symmetric basic
spaces with unit weight.

\section{The Dot Product $(\,\cdot\,)$}

Given two vectors $X, Y\in \C$, whose general form was given
in Eqn.\four,  we define the dot product
$X\cdot Y\in \C$, by the following two equations:
$$ \lb\A^{g_1}_{n_1}, \ldots \,,\A^{g_r}_{n_r}\rb
\cdot \lb\B^{h_1}_{m_1}, \ldots\,,
\B^{h_s}_{m_s}\rb \equiv  \lb\A^{g_1}_{n_1}, \ldots \,,\A^{g_r}_{n_r},
\B^{h_1}_{m_1},
\ldots \,,\B^{h_s}_{m_s}\rb\, , \eqn\edotone$$
$$ \Big(\sum_i a_i X_i\Big) \cdot \Big(\sum_j b_j Y_j\Big) \equiv \sum_{i,j}
a_i b_j\, X_i \cdot Y_j\, , \quad X_i, Y_j \in \C\, . \eqn\edottwo$$
It is manifest from this definition that $X\cdot Y \in \C$. For symmetric
basic spaces $\A^{g_i}_{n_i}$ and $\B^{g_i}_{n_i}$
with unit weight, the
spaces $X_1= \lb\A^{g_1}_{n_1}, \ldots \,,\A^{g_r}_{n_r}\rb$
and $Y_1= \lb\B^{h_1}_{m_1}, \ldots\,,
\B^{h_s}_{m_s}\rb$  have
the property that each inequivalent labeled surface appears with unit weight
(see the remarks below Eqn.\twopp). Since the right hand side of
Eqn.\edotone\ is also built as a symmetrized product of symmetric
basic spaces with unit weight, it follows
that in the dot product $X_1\cdot Y_1$
each inequivalent labeled surface appears with unit weight.

As defined, the dot product is manifestly associative.
It  also follows from the above definition, and Eqn.\three\ that
the dot product is graded commutative.  In summary
$$ X\cdot Y = (-)^{XY} Y\cdot X, \quad \quad
X \cdot (Y \cdot Z) = (X \cdot Y) \cdot Z, \quad \quad
X, Y, Z \in \C\,. \eqn\gradedcomm$$
We have therefore obtained the structure of a graded commutative
algebra on $\C$.

It is useful to have an alternative description of the dot product
$(X\cdot Y)$
for the case when $X$ and $Y$ are not themselves symmetrized products
but linear combinations theoreof. For this let us regard $X$ as a collection
of labeled surfaces $\Sigma_X$ with $N$ punctures appearing with weight
$w_{\Sigma_X}$,  and similarly $Y$
as a collection of labeled
surfaces $\Sigma_Y$ with $M$ punctures appearing with weight factor
$w_{\Sigma_Y}$.
We can then define $X\cdot Y$ as
the collection of surfaces of the form
$$ w_{\Sigma_X}w_{\Sigma_Y}\cdot
{1\over N! M!} \cdot \sum_{\sigma\in S_{N+M}}
 {\bf P}_\sigma \, (\Sigma_X \times \Sigma_Y\, ) \,. \eqn\simplemult$$
In constructing these class of terms we have relabeled the punctures
of the disjoint surfaces from one to $N+M$ and have symmetrized over
all possible assignment of labels to the punctures, dividing by the
symmetry factors $N!$ and $M!$ which take into account the original
symmetry of $X$ and $Y$. This definition is extended to more
general surfaces (say with variable numbers of punctures) by multilinearity.
In order to show that this definition is compatible with the earlier one
we must show how to derive \edotone\
from \simplemult. Using Eqn.\twopp\ and Eqn.\simplemult\ we get,
$$\eqalign{ \lb \A^{g_1}_{n_1}, \cdots , \A^{g_r}_{n_r}\rb
 \cdot\lb \B^{h_1}_{m_1}, \cdots , \B^{h_s}_{m_s}\rb &=
 {1\over n_1!\cdots n_r!}\cdot {1\over m_1!\cdots m_s!}
\sum_{\sigma'\in S_N}   \sum_{\sigma''\in S_M}  \sum_{\sigma\in
S_{N+M}} \cr
&\hskip-50pt\cdot {1\over N! M!} {\bf P}_\sigma {\bf P}_{\sigma'}
{\bf P}_{\sigma''}
(\A^{g_1}_{n_1}\times\cdots \times \A^{g_r}_{n_r}\times
 \B^{h_1}_{m_1}\times\cdots \times \B^{h_s}_{m_s})\,
\cr}\eqn\endhere$$
where $N= \sum n_i$ and $M = \sum m_i$,  $\sigma'$ is a permutation of
the labels $1$ to $N$ of the $\A$ spaces,
and  $\sigma''$ is a permutation of
the labels $1$ to $M$ of the $\B$ spaces. Due to the sum over the
$\sigma$ permutations of the labels $1$ to $N+M$, the $\sigma'$ and
$\sigma''$ permutations only contribute factors of $N!$ and $M!$
respectively. We therefore get
$$\eqalign{ \lb \A^{g_1}_{n_1}, \cdots , \A^{g_r}_{n_r}\rb
 \cdot\lb \B^{h_1}_{m_1}, \cdots , \B^{h_s}_{m_s}\rb &=
 {1\over n_1!\cdots n_r!}\cdot {1\over m_1!\cdots m_s!}
 \sum_{\sigma\in S_{N+M}} \cr
&\hskip-50pt\cdot {\bf P}_\sigma
(\A^{g_1}_{n_1}\times\cdots \times \A^{g_r}_{n_r}\times
 \B^{h_1}_{m_1}\times\cdots \times \B^{h_r}_{m_s})\, ,
\cr}\eqn\enhere$$
which, by Eqn.\twopp\ agrees with equation \edotone. This shows that
the two definitions of the dot product coincide.

It follows from Eqn.\simplemult\ that when we multiply two spaces
$X$ and $Y$ that only contain configurations  appearing with weights
equal to one, the dot product $(X\cdot Y)$ is made of different
configurations each appearing with unit weight (except in
degenerate cases). This is seen as follows. Each term we consider
in \simplemult\ would have $w_{\Sigma_X} = w_{\Sigma_Y} =1$, and we
get configurations with weight factor $1/N!M!$. Nevertheless, by symmetry
the spaces $X$ and
$Y$, contain respectively $N!$ copies of $\Sigma_X$ and $M!$ copies
os $\Sigma_Y$, with the punctures relabeled.
Each of these copies can be seen to
contribute an equal amount to $(X\cdot Y)$ by the rule expressed in
\simplemult. The number of different copies that can be combined
is $N!M!$. This cancels out the same weight factor appearing in the
denominator of \simplemult\ and results in every
single configuration produced with unit weight.

For our later
developments it will be convenient, though not strictly necessary,
to introduce a new element in the complex $\C$ which
will be a unit ${\bf 1}$ for the dot product. Thus, by definition,
$X\cdot {\bf 1}
= {\bf 1} \cdot X = X$,  for every $X\in \C$.
Intuitively, the unit can be thought to  represent the surface
with {\it zero} number of connected components and zero punctures.
As such, under the
dot product, which simply puts together the disconnected surfaces
of the spaces to be multiplied, multiplication by the unit has no effect.

The reader may note that the construction of the graded-commutative
associative algebra given here was done along the lines of the standard
construction in mathematics of free associative algebras starting from
a vector space $V$. In our case that vector space is the space of
basic subspaces of Riemann surfaces. As in the standard construction
one forms all tensor products  $V^{\otimes N}$ and adds them together to form
a complex. There is natural multiplication $V^{\otimes N}\times V^{\otimes M}
\to V^{\otimes (N+M)}$. Most of the work we had to do in our
construction was due to the necessity of working with symmetric spaces
throughout.

\chapter{The Batalin Vilkovisky Algebra}

In this section we begin by defining the $\Delta$ operator
and then turn to show that it squares to zero and that it is a second
order derivation of the dot product. This shows that we have
a Batalin-Vilkovisky
algebra [\getzler,\penkavaschwarz,\lianzuckerman].
We discuss how the antibracket is
recovered and explain the properties of the boundary operator $\partial$.
We also review the homomorphism to the BV algebra of string
functionals.
We conclude by showing how to extend the complex $\C$ to include elements
which are formal limits and have the interpretation of contractions
and Lie derivatives.

\section{The operator $\Delta$}

While our aim in this section is to give a definition for an
operator $\Delta$ acting on elements of $\C$ it will be convenient
to introduce an operator $\Delta_{i,j}$, with $i\not= j$,
 that will act on direct products
of basic spaces of surfaces (not necessarily symmetrized).

For any product space $\A$  ( a space of the form
$\A =\A^{g_1}_{n_1}\times\cdots\times \A^{g_r}_{n_r}$),
we define $\Delta_{ij}\A $ as ${\half}$ times the set of surfaces
obtained by twist
sewing the punctures $P_i$ and $P_j$ of every element in
$\A$. If $z_i$ and $z_j$ denote the local coordinates around those
punctures, twist sewing means sewing through the relation
$ z_i z_j = e^{i\theta}$ with $0\le \theta \le 2\pi$. It is clear
from the above definition that $\Delta_{ij} = \Delta_{ji}$. We extend the
definition of $\Delta_{ij}$ to linear combinations of product spaces
by taking it to be a linear operator.
The $\Delta_{ij}$ operation
reduces the number of punctures by two.
Therefore in the resulting surfaces the punctures must be
relabeled from $1$ to $N-2$.
This will be done preserving the ascending order of
the punctures.
If the two punctures to be sewn lie on the
same connected component, that connected component increases its genus
by one, and $\Delta_{ij} \A$ is made of disconnected surfaces
with $r$ components.
If the two punctures to be sewn lie on the
different connected components, those two connected components fuse to
give a single connected component. In this case $\Delta_{ij} \A$
has $r-1$ connected components.
The dimension of $\Delta_{ij} \A$ is dim$( \A)+1$,
with the twist angle $\theta$
parametrizing the extra dimension.
If $\{\A\}$ denotes the ordered basis of tangent vectors induced on
$\Delta_{ij}\A$
by the original basis of tangent vectors $[\A]$ of $\A$, we define the
orientation of the space
$\Delta_{ij} \A$ to be $[{\p\over\p\theta},
\{\A\}]$.\foot{Since
the phases of the local coordinates around the punctures of $\A$ are not
defined, the induced tangent vectors $[\A]$ of $\Delta_{ij}\A$ are
defined only up to addition of terms proportional to $\p/\p\theta$.
This ambiguity does not affect the definition of the orientation of
$\Delta_{ij}\A$ given above.}
If $\A$ represents
a product space with a total number of punctures less than or equal to
one, $\Delta_{ij}\A =0$.

For any $X= \lb \A^{g_1}_{n_1},\cdots,\A^{g_r}_{n_r}\rb \in \C$,
we now define
$$\Delta X  \equiv \Delta_{ij}X
={1\over n_1!\cdots n_r!}
\sum_{\sigma\in S_N} \Delta_{ij}
{\bf P}_\sigma  (\A^{g_1}_{n_1}\times\cdots
\times \A^{g_r}_{n_r})\eqn\deltadeff$$
where we made use of \twopp. The right hand side is actually independent
of the choice of $i$ and $j(\not= i)$ in $\Delta_{ij}$. This follows from
the symmetry of $X$. It is also clear that $\Delta X \in \C$.
The linearity of $\Delta_{ij}$ implies that $\Delta$  extends to
general elements of $\C$ as
a linear operator: $\Delta \sum _i a_i X_i \equiv
\sum_i a_i \Delta X_i\,$.

Let us make a comment about weight factors.
If $X= \lb \A^{g_1}_{n_1},\cdots,\A^{g_r}_{n_r}\rb$ is made of symmetric
basic spaces $\A^{g_i}_{n_i}$ with unit weight,
as mentioned below \normmult,
each different labeled
surface in $X$ has unit weight. Now we claim that the same holds for
$\Delta X$. This is so because
whenever we pick any two punctures to be
sewn in a surface in $X$,
the surface with those two punctures exchanged also appears with unit
weight in $X$ giving an identical contribution to $\Delta X$.
The explicit one-half factor in the
definition of $\Delta$ then restores unit weight.

We will define $\Delta$ to give zero on any element of $\C$ representing
a space of disconnected surfaces with a total number of punctures less than
or equal to one. Our definition of $\Delta$ does
not tell us how it acts on the unit
element ${\bf 1}$ of the algebra. We will define
$$\Delta \,{\bf 1} =0\,,\eqn\deltaonee$$
and this will actually be necessary for the consistency of the BV
algebra to be introduced later. It is also in accord with the intuitive
notion that the surface representing ${\bf 1}$,
with no connected components and no punctures,
does not admit a nontrivial action of $\Delta$.

\section{The BV Algebra Structure}

We shall now show that
the dot product $(\,\cdot\,)$ and the $\Delta$ operator satisfy the
properties defining a BV algebra.

\noindent 1. {\it The operator $\Delta$ is nilpotent:}
$$\Delta(\Delta X)= \Delta^2  X = 0 \,.\eqn\edeltasquare$$
Proof: This property holds, by definition
for the special element $X={\bf 1}$.
If $X$ denotes a subspace containing three or less punctures the
above property is clearly true, so we will now consider spaces whose
surfaces have
at least four punctures. Let $X$ be such a space and let us now
calculate $\Delta^2 X$ in two ways, first as  $\Delta_{12}\Delta_{12} X$,
and then as $\Delta_{12}\Delta_{34} X$. The independence of $\Delta$
from the choice of punctures guarantees that both evaluations must give
the same answer. We will show that
they differ by a sign, and therefore
the object is identically zero.

In calculating $\Delta_{12}\Delta_{12} X$
we first twist sew punctures $P_1$ and $P_2$ of every element of $X$ and
then relabel the punctures $P_3\cdots P_N$ as $P_1 \cdots P_{N-2}$ and
twist sew the new $P_1$ and $P_2$ punctures. This means that effectively
the second sewing operation is joining
the original $P_3$ and $P_4$ punctures.
In calculating $\Delta_{12}\Delta_{34} X$
we first twist sew punctures $P_3$ and $P_4$ of every element of $X$, and
then twist sew the punctures $P_1$ and $P_2$ which need no relabeling.
Let $\Sigma_X$ be an arbitrary element of $X$ appearing with some fixed
weight factor, and
representing a  surface with {\it labeled} punctures.
Let $\p/\p\theta_{12}$ be the tangent vector associated with the
sewing of the original punctures $P_1$ and $P_2$,
and $\p/\p\theta_{34}$ be the tangent vector associated with the sewing
of the original punctures $P_3$ and $P_4$ .
As explained above, the orientation of the space $\Delta_{12}\Delta_{12} X$
at the subspace
$\Delta_{12}(\Delta_{12}\Sigma_X)$ will contain the tangent
vectors $[ {\p\over \p\theta_{34}} ,  {\p\over \p\theta_{12} }, \{X\}]$
in this order. On the other hand
the orientation of the space $\Delta_{12}\Delta_{34} X$
at the subspace
$\Delta_{12}(\Delta_{34}\Sigma_X)$ will contain the tangent
vectors $[ {\p\over \p\theta_{12}} ,
{\p\over \p\theta_{34} }, \{X\}]$ in this
order. Thus the spaces $\Delta_{12}(\Delta_{12}\Sigma_X)$ and
$\Delta_{12}(\Delta_{34}\Sigma_X)$ are just the same space with opposite
orientation. This shows that the two ways of calculating $\Delta^2X$ give
answers that differ by a minus sign, and hence $\Delta^2 X =0$.

\noindent 2. {\it The operator $\Delta$ acts as a second order
super-derivation on the dot product:
$$\eqalign{
\Delta(X \cdot Y \cdot Z)\, =\, & \Delta (X\cdot Y) \cdot Z + (-)^X
X \cdot \Delta (Y \cdot Z) + (-)^{(X-1)Y} Y\cdot \Delta (X\cdot Z) \cr
& - \Delta X \cdot (Y \cdot Z) - (-)^X X\cdot (\Delta Y) \cdot Z
- (-)^{X+Y} X\cdot Y \cdot \Delta Z \cr}\eqn\secondone$$
where $X$, $Y$ and $Z$ are elements of $\C$ of definite dimensions. }

\noindent Proof:  By  the linearity of $\Delta$
and the multilinearity of the
dot product it is enough to consider the case when
$X$, $Y$ and $Z$ are spaces of the form
$$X = \lb \X_{l_1}^{f_1} , \cdots \X_{l_r}^{f_r} \rb\,,\quad
Y = \lb \Y_{m_1}^{g_1} , \cdots \Y_{m_s}^{g_s} \rb\,,\quad
Z = \lb \Z_{n_1}^{h_1} , \cdots \Z_{n_t}^{h_t} \rb\,.\quad \eqn\stup$$
Moreover, since $\C$ is spanned by symmetrized products of
symmetric basic
spaces with unit weight (see discussion at the end of sec.2.1)
there is no loss of generality in taking
all the basic spaces appearing here
to be symmetric and with unit weight.
Consider now the left hand side of Eqn.\secondone:
$$ \Delta (X\cdot Y \cdot Z ) = \Delta
\lb \X_{l_1}^{f_1} , \cdots \X_{l_r}^{f_r} ,
 \Y_{m_1}^{g_1} , \cdots \Y_{m_s}^{g_s},
\Z_{n_1}^{h_1} , \cdots \Z_{n_t}^{h_t} \rb\,.\quad\eqn\seetup$$
The right hand side of this equation contains a space of surfaces
which breaks naturally into six distinct classes. Let us denote
by $R_{XX}$ the subset of surfaces appearing in the right
hand side of Eqn.\seetup\ where the two punctures
sewn by $\Delta$ lie both on
surfaces belonging to $X$. Furthermore, let $R_{XY}$ denote
the subset of surfaces appearing in the right
hand side of Eqn.\seetup\ where one of the
two punctures sewn by $\Delta$ lies
on a surface belonging to $X$
and the other one on a surface belonging to $Y$.
We define $R_{YY}, R_{ZZ}$ and $R_{YZ}, R_{ZX}$ similarly.
The orientation of the $R$ spaces is taken to be induced by
the orientation of $\Delta (X\cdot Y \cdot Z)$, and therefore,
is given as  $\{ \p/\p\theta, [X],[Y],[Z]\}$.
This enables us to write \seetup\ as
$$ \Delta (X\cdot Y \cdot Z ) =  R_{XX} + R_{YY}+ R_{ZZ}+R_{XY}
+ R_{YZ} + R_{ZX}\,. \eqn\breakup$$
By the property of $\Delta$ giving unit weight to inequivalent
configurations when acting on a symmetrized product of symmetric
basic spaces with unit weight,
it follows that each inequivalent configuration appearing
in the right hand side of eqn.\seetup\ has unit weight.
Since \breakup\ is a disjoint breakup of this set, each configuration
appearing in any of the $R$ classes must appear with unit weight.

We now claim that
$$\Delta (X\cdot Y ) \cdot Z = R_{XX} + R_{XY} + R_{YY}\,,\eqn\aterm$$
As configurations it is clear that the set of surfaces on either side of
the equation is contained in the set of surfaces on the other side of the
equation. Moreover, the sets on the right hand side are all disjoint.
The only question is whether or not for each configuration the weight factors
agree. We have seen above that the right hand side contains each of its
inequivalent configurations with unit weight.  {}From our previous argument,
since  $(X\cdot Y)$ is a symmetrized product of
symmetric basic spaces of unit weight,
$\Delta (X\cdot Y)$ contains each inequivalent
configuration with unit weight. Since $Z$ is also of the same type, our
discussion of the dot product (below eqn. \enhere)
implies that $\Delta(X\cdot Y)\cdot Z$
must have each configuration with unit weight. Finally, the orientation
of the spaces on the two sides are identical. This proves Eqn.\aterm.

In an exactly identical manner, we can derive the following two equations:
$$(-)^X X\cdot \Delta (Y\cdot Z ) =
R_{YY} + R_{YZ} + R_{ZZ}\,,\eqn\bterm$$
$$(-)^{(X-1)Y}
Y\cdot \Delta (X\cdot Z ) = R_{XX} + R_{XZ} + R_{ZZ}\,,\eqn\cterm$$
where the extra sign factors appearing on the left hand sides of these
equations are required to take into account the necessary rearrangement
of the tangent vectors to bring them to the order $\{\p/\p\theta,
[X], [Y], [Z]\}$. Using similar arguments we can derive three more
useful equations:
$$\Delta X\cdot Y  \cdot Z = R_{XX} ,\eqn\dterm$$
$$ (-)^X X\cdot \Delta Y  \cdot Z = R_{YY}\,,\eqn\eterm$$
$$ (-)^{X+Y} X\cdot Y  \cdot \Delta Z = R_{ZZ} \,.\eqn\fterm$$
Eqn.\secondone\ follows immediately from Eqns.\breakup-\fterm. This concludes
our proof that $\Delta$ is a second order derivation of the dot product.

\section{Recovering the Antibracket}

It was shown in
Refs.[\wittenab,\getzler,\penkavaschwarz]
that given a graded
commutative and associative algebra with a second order
derivation which squares to zero, namely, a BV algebra,
one can reconstruct the standard BV antibracket.
In particular, one defines the anti-bracket $\{X, Y\}$ through the relation
$$ \{X,Y\} = (-)^X \Delta(X\cdot Y) + (-)^{X+1}(\Delta X)\cdot Y -
X\cdot (\Delta Y), \eqn\eantibracket $$
then the anti-bracket satisfies the usual BV algebra relations
$$\{ X ,  Y\}= -\,(-)^{( X+1)( Y+1)} \,\{ Y,   X \}\,,\eqn\comx$$
$$ (-)^{( X+1)( Z+1)} \Bigl\{ \{  X ,  Y\}\,  , \,  Z\Bigr\}\,+\,
\hbox{cyclic permutations of $X$, $Y$, $Z$} \, = 0\,.\eqn\jcbdntty$$
It also satisfies the following properties with respect to the dot product
$$ \{ X , Y\cdot Z\} = \{X, Y\}\cdot Z + (-)^{(X+1)Y} Y\cdot \{X,Z\}.
\eqn\eantidot $$

\noindent
\underbar{Geometrical Picture.}
As in our analysis of subsection 3.2, we can express the contribution
to $(-)^X\Delta(X\cdot Y)$ as sum of three different classes of surfaces.
We denote by $S_{XX}$ the subset of surfaces appearing in this term where
the two punctures sewn by $\Delta$ lie both on surfaces belonging to
$X$. We define $S_{YY}$ similarly. Furthermore, let $S_{XY}$ denote the
subset of surfaces appearing in this term where one of the two punctures
sewn by $\Delta$ lies on a surface belonging to $X$, and the other one
on a surface belonging to $Y$. The orientation of the $S$ spaces is taken
to be induced by the orientation of $(-)^X\Delta(X\cdot Y)$, and therefore,
is given as $\{[X], \p/\p\theta, [Y]\}$. This enables us to write,
$$ (-1)^X \Delta(X\cdot Y) = S_{XX} + S_{YY} + S_{XY}\, . \eqn\NEWONE $$
We also have, using arguments similar to the ones
used in subsection 3.2,
$$\eqalign{ (-)^X (\Delta X)\cdot Y = & S_{XX}\, , \cr
X \cdot (\Delta Y) = & S_{YY} \, .\cr
}
\eqn\NEWTWO
$$
Eqn.\eantibracket\ now gives:
$$ \{ X , Y \} = S_{XY}. \eqn\NEWTHREE$$
Thus the antibracket $\{X, Y\}$ has the following interpretation. If
$\Sigma_X$ denotes an element of $X$ and $\Sigma_Y$ denotes an element of
$Y$, then $\{X, Y\}$ consists of surfaces where one puncture of $\Sigma_X$
is sewn to one puncture of $\Sigma_Y$, and the final punctured surface is
symmetrized in all the external punctures.
The orientation of the resulting space is given by $\{[X], \p/\p\theta,
[Y]\}$.
We therefore recover
the definition of the antibracket given in Ref.[\senzwiebachtwo].

{}From the definition \eantibracket\ it also follows that $\{, \}$ is a
bilinear operator in $\C$
$$ \Big\{ \sum_i a_i X_i\, , \, \sum_j b_j Y_j \Big\}
= \sum_{i,j} a_i b_j \{ X_i \, , \, Y_j\} \, , \quad X_i, Y_j\in \C.
\eqn\ebracketone$$

\section{The Boundary Operator $\p$}

Besides the dot product $(\cdot)$, the $\Delta$ operator,  and the
antibracket $\{,\}$, there is another
useful operator that one can define in the complex $\C$. For any region
$\A\subset \wh\P^{(g_1,\ldots ,g_r)}_{(n_1,\ldots ,n_r)}$,
$\p\A$ will denote the boundary
of $\A$. The orientation of $\A$ induces an orientation on $\p\A$ as
usual. Given a point $p\in \p\A$, a set of basis vectors $[v_1, \cdots v_k]$
of $T_p(\p\A)$ defines the orientation of $\p\A$ if $[\,n,v_1, \cdots v_k\,]$,
with $n$ a basis vector of $T_p\A$ pointing outwards,\foot{To obtain
an outward vector one constructs a diffeomorphism between the neighborhood
of $p$ and a suitable half-space. The outward vector is the image under
the diffeomorphism of the standard normal to the half space (see, for
example [\spivak]).}
is the orientation of $\A$ at $p$.
The definition of $\p$ is extended over the whole complex $\C$
by treating it as a linear operator
$$\p \Big(\sum_i a_i X_i\Big) = \sum_i a_i \p X_i\, , \quad X_i
\in \C \eqn\epartialone $$
It is clear that acting on an element of $\C$, $\p$ gives another element
of $\C$. Also, $\p$ acts as an odd derivation of the
dot product
$$\p\, (X\cdot Y) =  (\p X\cdot Y)+
(-)^{X} \,(X\cdot\p Y)\, , \eqn\newxef$$
and anti-commutes with $\Delta$
$$\Delta\p X=-\p\Delta X\, . \eqn\edeltaboundary$$
Properties \newxef\ and \edeltaboundary\ follow from the geometric
definitions of $\p$, $(\cdot)$ and $\Delta$.
Finally, using
Eqns.\eantibracket, \newxef\ and \edeltaboundary\ we see that $\p$ acts
as an odd derivation of the anti-bracket:
$$ \p\, \{ X ,   Y\}  = \{ \p X ,   Y\} + (-)^{ X +1} \{  X ,
\p Y \} \, .\eqn\ebountimes$$

Following [\senzwiebachtwo] we now define an odd operator
$$\delta\equiv \p+\hbar\Delta\, ,\eqn\edefdelta$$
and verify that it squares to zero
$$\delta^2 = \bigl( \p + \hbar \Delta \bigr)^2 = \p^2 + \hbar (\p\Delta +
\Delta \p)  + \hbar^2 \Delta^2 =0\,.\eqn\newderiv$$
We can therefore define cohomology of $\delta$
in the complex $\C$. In the next section we shall see that
the string vertices that define
a string field theory can be naturally
associated to a cohomology class of $\delta$.

\section{ Representations
on the Space of Functions of String Fields}

For a string theory formulated
around any specific matter
conformal field theory with $c=26$, there is
a natural map from the subspaces of moduli spaces of punctured Riemann
surfaces to the space of functions of the string field. This map is
obtained via the objects $\bra{\Omega^{(k)g,n}}$, which are
$(\H^*)^n$ valued $(6g+2n-6)+k$ forms on $\wh\P_{g,n}$. Here $\H$
denotes the subspace of the Hilbert space of the combined
matter-ghost conformal field theory, annihilated by $b_0^-$ and
$L_0^-$, and $\H^*$ is the dual Hilbert space. (For the precise
definition of $\bra{\Omega^{(k)g,n}}$, see
Refs.[\zwiebachlong,\senzwiebachtwo].)
Given an element
$\lb \, \A_{g_1,n_1}^{(k_1)}, \ldots \A_{g_r,n_r}^{(k_r)}\rb $ of $\C$,
we define
$$ f\Big(\lb \, \A_{g_1,n_1}^{(k_1)},
\ldots \A_{g_r,n_r}^{(k_r)}\rb\Big) =  \prod_{i=1}^r {1\over  n_i!}
\int_{\A_{g_i,n_i}^{(k_i)}}\bra{\Omega^{(k_i)g_i,n_i}}\Psi\rangle_1\cdots
\ket{\Psi}_{n_i}\, . \eqn\dotrepone $$
Here, for convenience of writing, we have not included the string
field $\ket{\Psi}$ in the argument of $f$. This operation is extended
to the whole complex $\C$
by taking
$$ f\Big(\sum_i a_i X_i\Big) = \sum_i a_i f(X_i) \eqn\deffonc$$
where $a_i$ are any set of numbers and $X_i\in \C \,\, \forall i$.
The function $f(X)$ of a space $X$ of definite dimension, is
grasmann even if
the dimension of $X$ is even, and is grassmann odd if the dimension
of $X$ is odd.
The map $f$ is not defined  for spaces
containing zero, one, and two punctured spheres, as well as tori
without punctures. For higher genus
surfaces without punctures the map was given in Ref.[\senzwiebachtwo].
We now claim that the standard $\Delta$
and product operations in the space of string fields are related to the
corresponding operations in the moduli space in a simple manner:
$$\eqalign{
f\bigl( \Delta  X ) &= -\Delta  f( X) \, \cr
f( X \cdot  Y) &= f( X) \cdot f( Y)\, , \quad \quad X,Y\in \C\, .\cr
}\eqn\hreh$$
This is the homomorphism between the Riemann surface BV algebra, and
the BV algebra of string functionals.\foot{The minus sign of the first
equation could be eliminated if so
desired,  by changing the definition of either the geometrical
or the functional $\Delta$ operator.}
It follows from \eantibracket\ and the above equations that
$$f \bigl( \{  X ,  Y \} \bigr) = - \{\, f( X) , f( Y) \}\, .\eqn\abhom$$
The second equation of \hreh\ follows immediately from the definition
of $f$ in \dotrepone\ and the definition of the dot product in \edotone.
The derivation of the first
equation is somewhat more involved.
It was shown in Ref.[\senzwiebachtwo] that
$$\eqalign{
f\bigl( \Delta  \A ) &= -\Delta  f( \A) \, \cr
f \bigl( \{ \A ,\B \} \bigr) &= - \{\, f( \A ) ,f( \B) \}\, ,\cr}
\eqn\hrehp$$
for symmetric basic
spaces $\A , \B$. In order to establish the first
equation in \hreh\ an induction argument is useful. To begin with one
shows that the second equations in \hreh\ and \hrehp,
and \eantidot\ imply that
whenever $f \bigl( \{ X ,\A \} \bigr) = - \{\, f( X ) ,f( \A) \}\,$
holds for $\A$ a symmetric basic space and $X$ fixed, then
$f \bigl( \{ X\cdot \B ,\A \} \bigr) = - \{\, f( X\cdot \B ) ,f( \A) \}\,$
with $\B$ a symmetric basic space. This fact implies that
$$f \bigl( \{ X ,\A \} \bigr) = - \{\, f( X ) ,f( \A) \}\, , \eqn\newui$$
holds for {\it arbitrary} $X$ and $\A$ an arbitrary symmetric basic space.
Using the second equation in \hreh, first equation in \hrehp,
\newui\ and \eantibracket\ one can then show that whenever
$f\bigl( \Delta  X ) = -\Delta  f( X)$ holds for fixed $X$, then
$f\bigl( \Delta  (X\cdot\A) ) = -\Delta  f( X\cdot \A)$ holds with $\A$
a symmetric basic space. This fact, used in a simple induction argument,
implies the first Eqn. in \hreh. This concludes our verification of
the homomorphism.

\section{Contractions and Lie Derivatives of Spaces of Surfaces}

Given a vector field $\wh U$ in
$\wh\P^{(g_1,\cdots g_r)}_{(n_1,\cdots n_r)}$,
we define an operation that increases the degree of a
space of surfaces by one. Given a surface $\Sigma$ we let
$f_{*\wh U}^t\Sigma$  denote the surface obtained by following the integral
curve of the vector field $\wh U$ a parameter length $t$.
If $X$ is a subspace of
$\wh\P^{(g_1,\cdots g_r)}_{(n_1,\cdots n_r)}$
which is symmetric in all the
punctures, we define
$$ I_{\wh U}^u\, X \equiv  \Bigl\{ f_{*\wh U}^t\,X \,\,, \,
 t\in [0,u]\Bigr\}\, , \eqn\introdi$$
that is, the space of surfaces obtained by taking every element of $X$
and including in the resulting set all the surfaces obtained while following
the integral curves of $\wh U$ a parameter length $u$. The orientation of
$I_{\wh U}^u\, X $ will be defined by $[\wh U , [X]\,]$. Note that
this definition does not require that the vector field be defined over all
of $\wh\P^{(g_1,\cdots g_r)}_{(n_1,\cdots n_r)}$,
Given a space $X$ the vector field only needs to be
defined in a suitable neighborhood of $X$. We also
define
$$ L_{\wh U}^u\, X \equiv   f_{*\wh U}^u\,X \,- X \,, \,
 \eqn\introdl$$
which computes the difference between the space of surfaces we get by
following
the integral curve a parameter distance $u$ and the original space of
surfaces. It follows
from the definitions given above that
$$ \partial\, I_{\wh U}^u\, X  \,= \,-\,  I_{\wh U}^u\,\partial  X
+  \, L_{\widehat U}^u X \,,\eqn\partiii$$
since the boundary operator picks two types of contributions, one from
the boundary of $X$ (with a minus sign because both $\partial$ and $I$ are
odd), and the other from the endpoints of the displacement along the
integral curves of $\wh U$.

We now extend our complex $\C$ by including in it the following formal limits
$$\eqalign{ i_{\widehat U} X &\equiv \lim_{u\to 0}\,\,
{1\over u}\,\, I_{\widehat U}^u X  \, \,, \cr
 \L_{\widehat U} X &\equiv \lim_{u\to 0}\,\, {1\over u}\,\,
 L_{\widehat U}^u X \, .\cr} \eqn\firstadd$$
This defines the linear operators $i_{\widehat U}$ and
$\L_{\widehat U}$ in the complex $\C$. We can now
use the definitions given in eqns.(4.6) and (4.8)
of Ref.[\hatazwiebach] to verify that
$$\eqalign{
 \int_{i_{\widehat U  }X}
\Omega &= \int_{X} i_{\widehat U}\Omega \,,\cr
\int_{\L_{\widehat U  }X} \Omega &=  \int_{X} \L_{\widehat U}\Omega \,,
\cr}\eqn\thirdadd$$
where, in the right hand sides $i_{\widehat U}\Omega$ and
$\L_{\widehat U}\Omega$ denote respectively
the contraction operation and Lie derivative
on the canonical forms $\Omega$ appearing in Eqn.\dotrepone.
We now impose the following identification on the new elements
$i_{\wh U} X$ and $\L_{\wh U} X$ of $\C$:
$$\eqalign{
i_{\widehat U_1 +\widehat U_2} X &= i_{\widehat U_1} X +
i_{\widehat U_2} X \,, \cr
\L_{\widehat U_1 +\widehat U_2} X &= \L_{\widehat U_1} X +
\L_{\widehat U_2} X \, .\cr} \eqn\fifthadd$$
These identifications are compatible with Eqn.\thirdadd\ since the
contraction of forms
$i_{\wh U}$  and  the Lie derivative of forms $\L_{\wh U}$ are
both linear on
the vector field argument $\wh U$.\foot{This means that
$(i_{\wh U_1 + \wh U_2}X - i_{\wh U_1}X - i_{\wh U_2}X)$ is in the
kernel of the homomorphism $f$ defined in eq.\dotrepone. Since
ultimately we are interested in applying this formalism to string
theory, we do not lose anything by defining this difference to be
zero in $\C$ itself.}
In terms of the new objects Eqn.\partiii\ implies that
$$ \partial\, i_{\wh U}^u\, X  \,= \,-\,  i_{\wh U}^u\,\partial  X
+  \, \L_{\wh U}X \,.\eqn\piii$$

\chapter{Closed String Vertices as Cohomology in $\C$ and a Lie Algebra}

In the present section we will begin by showing that the element
$e^{\V/\hbar} \in \C$, where $\V$
is the sum of the closed string field theory
vertices, is annihilated by the odd operator $\delta$ introduced
in sect.3.3. Since $\delta$ squares to zero, $e^{\V/\hbar}$ is a candidate
for a cohomology class.  We then reconsider the work of Ref.[\hatazwiebach]
and show that the difference between $e^{\V/\hbar}$ and
$e^{{\V'}/\hbar}$, where
$\V$ and $\V'$ are consistent string vertices, is a $\delta$-trivial term.
We conclude by discussing a background independent Lie algebra that
is constructed using the cohomology class $\V$
and is isomorphic to a subalgebra of the
string field theory gauge algebra.

\section{Closed String Vertices as a Cohomology Class}

The vertices of a string field theory can be associated with
$(6g+2n-6)$ dimentional subspaces $\V_{g,n}$ of $\wh\P_{g,n}$,
satisfying the recursion relations[\zwiebachlong]:
$$\p\V_{g,n} =- \half\sum_{g_1+g_2=g\atop n_1+n_2 = n+2}
\hskip-6pt\bigl\{ \V_{g_1,n_1}\, , \, \V_{g_2, n_2}\bigr\}\, -\,
\Delta \V_{g-1, n+2}\, .\eqn\propvgn$$
Let us now define
$$\V \equiv  \sum_{ g,n}\hbar^g \V_{g,n}  \quad\hbox{with}\quad
\cases{n\geq 3 \,\,\hbox{for}\,\,g=0 ,\cr
n\geq 1\,\,\hbox{for}\,\,
g=1,\cr n\geq 0 \,\, \hbox{for}\,\, g\ge 2\,.}\eqn\dfnvsp$$
It then follows from \propvgn\ that the recursion relations can be written as
$$\p \V   + \hbar \Delta \V + \half \{ \V , \V \} = 0 \, .\eqn\recrelnew$$

We shall now show that a $\V$ satisfying this equation defines a
cohomology element of $\delta$.
We define the
exponential function  of an even element $X\in\C$ by the usual
power series
$$\exp( X) \equiv {\bf 1}
 +  X + {1\over 2}  X\cdot  X + {1\over 3!}  X\cdot X\cdot X
+ \cdots \,.\eqn\expdef$$
It follows from Eqn.\newxef\ that
$$\p \,\bigl[ \,\exp ( X)\,\bigr] = \p  X \cdot\exp( X)\, ,\eqn\diffexp$$
Moreover, using Eqn.\eantibracket\ we find
$$ \Delta \exp ( X) = \bigl( \Delta  X + \half
\{  X ,  X\} \bigr) \exp ( X)\,.\eqn\gtmq$$
{}From the last two equations we get,
$$ \delta\exp( X) = \bigl( \p + \hbar \Delta \bigr)
\exp ( X) = \bigl(\p  X + \hbar \Delta  X +
\half\hbar
\{  X ,  X\} \bigr) \exp ( X)\,.\eqn\gtmqx$$
Making use of Eqn.\gtmqx\ we see that we can now write the recursion
relations \recrelnew\ in the simple form
$$ \delta \exp (\V/\hbar) = 0\,.\eqn\sform$$
Thus $\exp(\V/\hbar)$ defines a cohomology element of $\delta$.
It is clear that $\exp (\V/\hbar)$
is not $\delta$ trivial since the expansion begins with
${\bf 1}$, and a term of the form $\delta Y \equiv \p Y + \hbar \Delta Y$
can never contain a term proportional to ${\bf 1}$. This result, however,
is not very interesting, since even for $\V=0$, $exp(\V/\hbar)
={\bf 1}$ will define a non-trivial element of the cohomology. A more
interesting fact is that even $\exp(\V/\hbar)-1$, which is $\delta$
closed by virtue of Eqn.\sform,  is not $\delta$ trivial.
Triviality would require that
$$\hbar^{-1}\,\V + \half\hbar^{-2}\, \V \cdot \V +\cdots
=(\partial + \hbar\Delta ) \{ X \}\,,\eqn\hir$$
More explicitly, this equation begins as
$$\hbar^{-1}\,\V_{0,3} + \cdots
= (\partial + \hbar\Delta ) \{ X  \}\,.\eqn\isittr$$
If we are to obtain the zero-dimensional space
$\V_{0,3}$ from the right hand side
it cannot be from $\Delta$ since $\Delta$ always adds one
dimension. Thus it must be from
$\partial$. But $\V_{0,3}$ cannot be written as $\partial X$ for any $X$.
This shows that $(\exp(\V/\hbar) -1)$ is not trivial.
This is the precise statement we have in mind when we state that the
string vertices define a cohomology class of $\delta$.

We also note that given a set of closed string vertices satisfying the
recursion relations  \propvgn, we can introduce a nilpotent operator
$\delta_\V$ through the relation
$$\delta_\V = \delta + \{\V, \, \}, \eqn\defdeltav$$
which has the property
$$\delta_\V \{ \A , \B \} =
\{ \delta_\V \A , \B \} +
(-)^{\A+1} \{ \A , \delta_\V \B \}\, , \eqn\delvone$$
and,
$$ \delta \left( \A e^{\V/\hbar} \right)  = (\delta_\V \A )
e^{\V/\hbar}\, , \eqn\delvtwo$$
for arbitrary subspaces $\A$, $\B$. This $\delta_\V$ operator will
be useful later on.

\section{Changing the Closed String Vertices}

In string field theory, the vertices $\V_{g,n}$ defining the $\delta$
cohomology class $e^{\V/\hbar}$
are not unique. The simplest choice for the
vertices appears to be that determined by the minimal area problem
[\zwiebachlong]
and a simple example of a family of consistent closed string vertices
is given by the simple deformation of attaching stubs to the vertices.
In Ref.[\hatazwiebach] the general situation when we have a parametrized
family of consistent string vertices $\V_{g,n}(u)$ was studied. It
was shown that for infinitesimally close string vertices, the
resulting  string field theories are related by an infinitesimal
(though nonlinear) string field redefinition. This redefinition
respects the antibracket and its explicit form was found.

With the insight that we have obtained into the string vertices, it is
natural to expect that $e^{\V(u)/\hbar}$
actually represents the same cohomology
class of $\delta$ for all values of $u$. If so, we should be able
to establish a relation of the form
$${d\over du} e^{\V(u)} = \delta\, \bigl(\,\chi(u)\,\bigr)\eqn\desiret$$
for some $\chi$. We shall now show that this is indeed the case.

The fact that we have a family of string vertices
$\V(u)$ implies that the recursion relations are satisfied for each
value of the parameter $u$
$$\partial\V (u) = -\half\,\{ \V (u),\V (u)\} -
\hbar\Delta\V (u).\eqn\putname$$
A geometrical fact established
in [\hatazwiebach] was the existence,
for each moduli space, of a vector field $\wh U$
such that
$$ f_{*\wh U}^{u_0} \V(u) = \V(u+u_0) \, .\eqn\sewcompt$$
This vector $\widehat U$ was constructed recursively.\foot{In
Ref.[\hatazwiebach] the vector $\widehat U$ satisfied the further
requirement that the deformation of each $U(1)$ fiber in $\p\V$ was
defined by the deformation of the constituent surface(s) appearing in
the right hand side of \putname\ and representing the basepoint of
the fiber. This additional requirement is not
necessary for the present proof. In Ref.[\hatazwiebach] this extra
requirement implied that integrals that had to be equal were so
by the manifest equality of their integrands.}

Consider now infinitesimal
variations $du$ and define
$$ \W (u) \equiv  \{ \V(u'): \, \, u'\in
[u, u+du] \}
= I_{\wh U}^{du}\, \V(u) = du\, i^{du}_{\wh U} \V(u)
\, ,\eqn\NEWEQN$$
where, by definition, the orientation of a space $\{\A(t):\, t\in \D\}$ is
given by the ordering $\{\p/\p t, [\A(t)]\}$ of the tangent vectors.
Using Eqn.\partiii, we find
$$\partial \W = \V(u+du) - \V(u) -I_{\wh U}^{du} \partial \V(u)\,,\eqn\name$$
where explicitly
$$\eqalign{
I_{\wh U}^{du} \partial \V(u) &= \{
 \partial \V (u'): \,\,
u'\in [u, u+du]\}\,,\cr
&= \Big\{ -\half\{ \V(u') , \V(u') \} - \hbar\Delta \V(u'): \,\,
 u'\in [u, u+du]\Big\}\,, \cr
&= \Big\{ - \half\{ \V(u') , \V(u') \}: \, \,
u'\in [u, u+du] \Big\}
+\hbar \Delta \W(u)\quad\quad\,\,.
\cr}\eqn\dffe$$

Consider now the region
$R^{g_1,g_2}_{n_1,n_2}(u,v) \in \wh\P^{g_1+g_2}_{n_1+n_2-2}$,
corresponding to the collection of surfaces
$\{ \V_{g_1, n_1}(u) ,\V_{g_2, n_2}(v) \}$ obtained twist-sewing
string vertices for {\it fixed nearby}
values of $u$ and $v$.\foot{Since
$R^{g_1,g_2}_{n_1,n_2}(u,u)$ is assumed to be a submanifold of
$\wh\P^{g_1+g_2}_{n_1+n_2-2}$,
$R^{g_1,g_2}_{n_1,n_2}(u,v)$ will also be a submanifold of
$\wh\P^{g_1+g_2}_{n_1+n_2-2}$ for
$v$ sufficiently close to $u$. It is not clear, however, that
a disjoint union of the various $R^{g_1,g_2}_{n_1,n_2}(u,v)$,
for different values of $u$ and $v$ form a submanifold of
$\wh\P^{g_1+g_2}_{n_1+n_2-2}$.}
Let us now introduce two vector fields $ U_1(u,v)$ and $ U_2(u,v)$
on $R^{g_1,g_2}_{n_1,n_2}(u,v)$ as follows.
$$
f^t_{* U_1(u,v)}\{ \V_{g_1, n_1}(u), \V_{g_2, n_2}(v) \}
= \{ \V_{g_1, n_1}(u+t), \V_{g_2, n_2}(v)\} + \O(t^2)\,,
\eqn\equationone $$
$$
f^t_{* U_2(u,v)}\{ \V_{g_1, n_1}(u), \V_{g_2, n_2}(v) \}
= \{ \V_{g_1, n_1}(u), \V_{g_2, n_2}(v+t)\} + \O(t^2)\, .
\eqn\equationtwo $$
These equations do not determine the vector fields
$U_1(u,v)$ and $U_2(u,v)$ uniquely. One way to fix a choice is
to demand that
$$f^t_{* U_1(u,v)}\{ \Sigma_1\in \V_{g_1, n_1}(u)\, ,
\,\Sigma_2\in \V_{g_2, n_2}(v) \}
= \{f^t_{*\wh U} \Sigma_1\, , \,\Sigma_2\} + \O(t^2)\,,
\eqn\equationonep$$
and similarly for $U_2 (u,v)$,
where $\wh U$ is the vector field appearing in \sewcompt. This defines
the map of $U(1)$ classes arising from twist sewing.
A map of the surfaces themselves is obtained by fixing arbitrarily the
phases around the punctures to be sewn, as discussed in sect.4.7 of
Ref.[\hatazwiebach].

We will now single out two special vector fields
$$\wh U_1 \equiv  U_1 (u,u) \,,\quad \wh U_2 \equiv  U_2 (u,u)
\, ,\eqn\spvect$$
defined on $R^{g_1,g_2}_{n_1,n_2}(u,u)$, and we extend them arbitrarily
but smoothly over
some neighborhood of $R^{g_1,g_2}_{n_1,n_2}(u,u)$. We will still denote by
$\wh U_1$ and $\wh U_2$ the
extended vector fields. It follows from
Eqns. \equationone\ and \equationtwo, together with
Eqns.\introdi\ and \firstadd\ that
$$\eqalign{
I_{\wh U_1}^{du} \{\V(u), \V(u)\} \equiv
du \, i_{\wh U_1}\{ \V(u) , \V(u) \} &=
\Big\{\{ \V(u') , \V(u) \}:  \, \,
 u'\in [u, u+du]\Big\}\cr
&=\{ \W(u) , \V(u) \}\,,\cr} \eqn\youone$$
$$\eqalign{
I_{\wh U_2}^{du} \{\V(u), \V(u)\} \equiv
du \, i_{\wh U_2}\{ \V(u) , \V(u) \} &=
\Big\{\{ \V(u) , \V(u') \}: \, \,
 u'\in [u, u+du]\Big\}\cr
&=-\{ \V(u) , \W(u) \}\,,\cr} \eqn\youtwo$$
The minus sign on the right hand side of the last equation can be traced to
different locations of the tangent vector $\p/\p u$ on the two sides
of the equation.

We now consider the deformation of $\{ \V(u) , \V(u) \}$ by
the vector $\wh U_1 + \wh U_2$.
The quantity $f^t_{*(\wh U_1 + \wh U_2)}\{\V(u), \V(u)\}$ corresponds to
following the integral curve of the vector field $\wh U_1 + \wh U_2$ for
a parameter distance $t$. Since the vector fields $\wh U_i$ have been
defined in a neighborhood of $R(u,u)= \cup R^{g_1,g_2}_{n_1,n_2}(u,u)$,
this is a well defined operation
for small enough $t$.
Moreover, for sufficiently small $t$,
$f^t_{*(\wh U_1 + \wh U_2)}\{\V(u), \V(u)\}$ can be obtained
by first following the integral curve of the vector field
$\wh U_1$ over a distance $t$, and then, starting at that deformed surface,
following the integral curve of the vector field $\wh U_2$
over a distance $t$.
This is correct to order $\O (t^2)$ since the vector fields $\wh U_1$ and
$\wh U_2$ are smooth.  We therefore have
$$\eqalign{
f^t_{*(\wh U_1 + \wh U_2)}\{ \V(u), \V(u)\}
&= f^t_{*\wh U_2}
(f^t_{*\wh U_1} \{\V(u), \V(u)\}) + \O(t^2)\, , \cr
&= f^t_{*\wh U_2}\{\V(u+t), \V(u)\} + \O(t^2) \,, \cr
}\eqn\notyet $$
The $\wh U_2$ in the second line
refers to the smooth extension of the vector $U_2(u,u)$.
This differs from the vector $U_2(u+t, u)$ by a term of order $t$.
Since we are ignoring the order $t^2$ terms in our analysis, we can
replace the $\wh U_2$ in the above equation by $U_2(u+t, t)$, and
then, by virtue of Eqn.\equationtwo, the right hand side of Eqn.\notyet\
can be replaced by $\{\V(u+t), \V(u+t)\}$. This gives,
$$
f^t_{*(\wh U_1 + \wh U_2)}\{ \V(u), \V(u)\}
= \{\V(u+t), \V(u+t)\} + \O(t^2)\, . \eqn\equationfour$$

{}From Eqns.\introdi, \firstadd, and \equationfour\
it now follows that
$$\Big\{ \{ \V(u') , \V(u') \}: \, \,
 u'\in [u, u+du] \Big\}
= I_{\wh U_1 + \wh U_2}^{du} \{ \V(u), \V(u) \}
=  du \, i_{\wh U_1 + \wh U_2}\{ \V(u) , \V(u) \}\, , \eqn\onex$$
and therefore
$$\eqalign{
\Big\{ \{ \V(u') , \V(u') \}: \, \,
 u'\in [u, u+du] \Big\}
&= du \,( i_{\wh U_1} + i_{\wh U_2})\, \{ \V(u) , \V(u) \} \cr
&= 2 \,\{ \W(u) , \V(u) \}\, ,\cr }
\eqn\youonetwo$$
where use was made of \fifthadd, \youone\ and \youtwo.
Back to \dffe\ we have
$$I_{\wh U}^{du} \partial \V(u)
= \{ \V(u) , \W(u) \} + \hbar\Delta \W(u)\, , \eqn\newfdde$$
and using \name\ and  \defdeltav\ we get
$$\eqalign{
\V(u+du) - \V(u) &= \partial \W + \{
\V(u) , \W(u) \} + \hbar\Delta \W(u) \, ,\cr
&= \delta_\V \W(u)\, . \cr}\eqn\nameless$$
Using Eqn.\delvtwo, the above equation can be rewritten as,
$$ \exp(\V(u+du)/\hbar) - \exp(\V(u)/\hbar) = {1\over \hbar} \delta
\big(\W\exp(\V/\hbar)\big)\, .
\eqn\ehzwfour $$
This shows that $\exp(\V(u+ du)/\hbar)$ and $\exp(\V(u)/\hbar)$
belong to the same cohomology class of $\delta$.

\section{Gauge Transformations}

Here we wish to note the existence
of a background independent  Lie algebra intimately
connected to gauge transformations. It is not quite the usual gauge
transformations, which can only be written in a background dependent
way, but it is isomorphic to a subalgebra of the full gauge algebra, and
could be  closely related to
the underlying gauge symmetry of a manifestly background
independent string field theory.

In ref.[\senzwiebachgauge], the space of gauge parameters was
identified with the space of hamiltonian fuctions $\Lambda$
of the string field,
with the identification $\Lambda \equiv \Lambda + \Delta_S F$. Here
$\Delta_S = \Delta + {1\over \hbar} \{ S , \cdot \}$
is the delta operator associated to the measure
$d\mu \exp (2S/\hbar)$ [\schwarz]. We then
had a Lie algebra ${\cal L}_{GT}$ of gauge transformations defined
by a bracket $[\cdot , \cdot ]$
$$\eqalign{
[ \Lambda_1 , \Lambda_2 ] & \equiv
\{ \Lambda_1 ,\Delta_S \Lambda_2 \} , \cr
&= (-)^{\Lambda_1} \{ \Delta_S \Lambda_1 , \Lambda_2 \} , \cr
&= \half \left(  \{ \Lambda_1 , \Delta_S \Lambda_2 \}
- (-)^{\Lambda_1 \Lambda_2 }  \{ \Lambda_2 , \Delta_S \Lambda_1\}
\right)\, . \cr} \eqn\qgt$$
Note that since $\Delta_S$ acts as an odd derivation of the
anti-bracket [\senzwiebachgauge],
the differences
between the various right hand sides of the above equation are all
$\Delta_S$ exact, and hence vanish in the space of gauge parameters.
Furthermore, if we add a $\Delta_S$ exact quantity to either $\Lambda_1$
or $\Lambda_2$, then by virtue of the nilpotence of $\Delta_S$,
$[\Lambda_1, \Lambda_2]$ defined in the above equation changes by a
$\Delta_S$ exact quantity.

The homomorphism between the Riemann surface BV algebra and the string field
BV algebra is easily shown to imply that [\senzwiebachtwo]
for any $X\in \C$
$$f (\delta_\V X ) = -\Delta_S \, f(X), \eqn\standhom$$
where $\delta_\V = \p + \hbar\Delta + \{ \V , \cdot \}$. This suggests a way
to obtain a Lie algebra at the level of Riemann surfaces by a construction
similar to that given above.
At the level of Riemann surfaces we now form the
space $\C_\V$ of equivalence classes $X \approx X + \delta_\V Y$
in  $\C$
and in this space define
a Lie algebra ${\cal L}_{RS}$
$$\eqalign{
[ X_1 , X_2 ] & \equiv
\{ X_1 ,\delta_\V X_2 \} , \cr
&= (-)^{X_1} \{ \delta_\V X_1 , X_2 \} , \cr
&= \half \left(  \{ X_1 , \delta_\V X_2 \}
- (-)^{X_1 X_2 }  \{ X_2 , \delta_\V X_1\}
\right)\, . \cr} \eqn\newlie$$
As before, the difference between various lines of the above equation
vanishes in $\C_\V$, and, furthermore, $[X_1, X_2]$ depends only on
the representative classes of $X_1$ and $X_2$ in $\C_\V$.
The Jacobi identity
$$ (-)^{X_1 X_3} \big[ [X_1, X_2], X_3\big] \, + \,
\hbox{cyclic permutations of $X_1$, $X_2$, $X_3$ } \, =0\,,\eqn\liejacobi$$
can be verified using the following three equations:
$$\eqalign{
 (-)^{X_1X_3} [ [ X_1, X_2 ] , X_3 ] &=
(-)^{X_3} (-)^{ (X_1 + 1) X_3 } \{ \{ X_1 , \delta_\V X_2\},
\delta_\V X_3 \}\, \cr
 (-)^{X_2X_1} [ [ X_2, X_3 ] , X_1 ] &=
(-)^{X_3} (-)^{ (X_1 + 1) X_2 } \{ \{ \delta_\V X_2,\delta_\V X_3\},
  X_1 \}\,\cr
 (-)^{ X_2 X_3} [ [  X_3,  X_1 ] ,  X_2 ] &=
(-)^{X_3} (-)^{X_2 X_3 } \{ \{ \delta_\V X_3,  X_1\},
\delta_\V  X_2 \}\, , \cr}\eqn\jacobithree $$
and the Jacobi identity \jcbdntty\ for the BV anti-bracket.

By virtue of \standhom\ $f$ induces a well defined
map between the Riemann surface Lie algebra $\L_{RS}$ and
the Lie algebra of gauge transformations $\L_{GT}$.
The map is actually a homomorphism. Indeed
$$\eqalign{
[ f( X_1) , f( X_2) ] &= \{ f(X_1) , \Delta_S f( X_2) \} \, , \cr
&= - \{ f( X_1) ,  f(\delta_\V X_2) \} \, , \cr
&= f \left( \{  X_1 ,  \delta_\V X_2 \}\right) \, , \cr
&=  f \left( [  X_1 ,  X_2 ] \right) \, , \cr} \eqn\prhmalg$$
where use was made of Eqns.\standhom\ and \abhom.
This homomorphism is clearly not an isomorphism.
The hamiltonians of the form $\bra{\omega_{12}}\Lambda\ket{\Psi}$,
that generate the usual gauge transformations with parameter
$\ket{\Lambda}$, are missing. This is so because
the map $f$ is not defined for one-punctured spheres
and we therefore cannot get hamiltonians linear on the string field.
While linear hamiltonians can be obtained from higher genus
surfaces with one puncture, it seems clear they
are not general enough to reproduce all possible
standard gauge transformations.
In particular, since all one point functions conserve momentum (and
other quantum numbers) one cannot  get a $\ket{\Lambda}$
with non-zero momentum.
We therefore expect that $f(\L_{RS})$ is  only a subalgebra $\wh\L_{GT}$ of
$\L_{GT}$. By the standard property of homomorphisms $\wh\L_{GT}$ is
isomorphic to the quotient Lie algebra $\L_{RS}/\hbox{Ker}(f)$,
where Ker($f$) is the ideal of $\L_{RS}$ generated by the elements that
map to zero.

Now, as has been emphasized before, the choice of $\V$ is not unique, but
there are whole families of vertices that satisfy the recursion relations
\propvgn\ and hence can be used to construct a closed string field theory.
We shall now argue that the Lie algebra of gauge transformations defined
above is independent of the choice of $\V$.
Consider another lie algebra $\L'$ defined on the
complex $\C_{\V'}$ of equivalence classes $ X \approx  X + \delta_{\V'}
 Y$, where $\V' = \V + \delta_\V \W $. In this algebra
$$[  X_1 ,  X_2 ]' = \{  X_1 , \delta_{\V'}  X_2 \}\, , \eqn\bartonone $$
where
$$ \delta_{\V'} = \delta _\V + \{ \delta_\V \W , \cdot \}\, .
\eqn\bartontwo$$
Consider the map $m : \C \to \C$ defined
as
$$ m( X) =  X + \{  X , \W \}\, , \eqn\bartonthree$$
where we consider $\W$ to be small. This map gives
an automorphism of the dot algebra
$$ m ( X \cdot  Y ) = m( X) \cdot m( Y)\, , \eqn\bartonfour $$
as one verifies using \eantidot. It is not, however,
an automorphism of the BV algebra.
Moreover using the Jacobi identity of the antibracket
one sees that
$$ \{ m( X) , m( Y) \} = m ( \{  X ,  Y \} )\, . \eqn\bartonfive$$
We now verify that
$$ \eqalign{
\delta_{\V'}  m( X) & = \delta_\V m( X) + \{ \delta_\V \W,m( X)\}\, ,\cr
&= \delta_\V  X + \delta_\V \{  X,\W \} + \{ \delta_\V\W, X \}\,, \cr
& = \delta_\V  X + \{ \delta_\V X ,\W \}\, , \cr }
\eqn\bartonsix$$
and therefore
$$ \delta_{\V'}  m( X) = m ( \delta_\V  X )\, . \eqn\nice$$
This means that the map $m$ induces a map from
$\C_\V$ to $\C_{\V'}$, since zero elements are mapped to
zero elements.
Since the map $m$ is invertible,
the isomorphism between $\L$ and $\L'$ is established if we show that
$$ [ m( X_1) , m( X_2) ]' = m \left( [ X_1 ,  X_2 ] \right)\, .
\eqn\bartonseven $$
This is now quite simple:
$$\eqalign{
[ m( X_1) , m( X_2) ]'  &= \{ m( X_1) , \delta_{\V'}m( X_2) \}\, , \cr
&= \{ m( X_1) , m(\delta_\V X_2 ) \} \, , \cr
&= m \left( \{  X_1 , \delta_\V  X_2 \} \right) \, ,\cr
&= m \left( [  X_1 ,  X_2 ] \right)\, . \cr}
\eqn\bartoneight $$
And this concludes the proof that the Lie algebra $\L_{RS}$
is universal in the sense that it does not depend on the specific
choice of consistent string vertices.

\ack We would like to thank S. Mukhi and E. Witten
for useful discussions. A. Sen would like to thank the hospitality
of Center for Theoretical Physics, M.I.T., where part of this
work was done.
B. Zwiebach would like to thank the hospitality of the Aspen Center
where this work was completed.

\singlespace
\refout
\end